\documentclass[a4paper,11pt]{article}
\pdfoutput=1 
\newcommand{\comm}[1]{}
\usepackage{xcolor}
\usepackage{pagecolor}
\usepackage{jinstpub} 
\usepackage{lineno}
\linenumbers
\usepackage{siunitx}

\title{\boldmath System Integration of ATLAS ITK Pixel DCS ASICs}

\author[a]{A. Qamesh,}
\author[a]{R. Ahmad,}
\author[a]{D. Ecker,}
\author[c]{T. Fischer,}
\author[b]{M. Karagounis,}
\author[a]{P. Kind,}
\author[a]{S. Kersten,}
\author[c]{T. Krawutschke,}
\author[b]{L. Schreiter,}
\author[a]{C. Zeitnitz,}

\affiliation[a]{University of Wuppertal,\\Fakultät 4, \\Gaussstr. 20, 42119 Wuppertal, Germany}
\affiliation[b]{Fachhochschule Dortmund - University of Applied Sciences and Arts ,\\Sonnenstr. 96, 44139 Dortmund, Germany}
\affiliation[c]{Technische Hochschule Köln,\\Ubierring 40, 50678 Köln, Germany}
\emailAdd{ahmed.qamesh@cern.ch}

\abstract{During the ATLAS phase II upgrade, the tracking system of the ATLAS experiment will be replaced by an all-silicon detector called the inner tracker (ITK) with a pixel detector as the most inner part. The monitoring data of the new system will be aggregated from an on-detector ASIC called Monitoring Of Pixel System (MOPS) and sent to the Detector Control System(DCS) using a new interface called MOPS-HUB. The hardware components of the MOPS-HUB, firmware specifications for the FPGA of MOPS-HUB and its integration plan will be presented. In addition, an irradiation plan for the new system will be introduced.}

\keywords{Front-end electronics for detector readout, Detector control systems (detector and experiment monitoring and slow-control systems, architecture, hardware, algorithms, databases), Digital electronic circuits, Detector control systems, Digital signal processing (DSP)}

\collaboration[c]{on behalf of ATLAS ITK collaboration}

\proceeding{Topical Workshop on Electronics for Particle Physics- TWEPP$2022$\\
  $19-23$ September\\
  Bergen, Norway}

\begin{document}
\nolinenumbers
\maketitle
\flushbottom
\section{Introduction}
\label{sec:introduction}
The ATLAS experiment will get a new ITK during the phase II upgrade, the innermost part will be a pixel detector. The new ATLAS pixel detector will use a serial powering scheme to reduce the number of services inside the detector volume~\cite{a,b}. Therefore, a new DCS is being developed at the University of Wuppertal to fulfill the control and monitoring requirements of the new pixel detector~\cite{c}. The new DCS will have an on-detector ASIC called MOPS to monitor the voltages and temperatures of the detector modules and other sub-detector components~\cite{Walsemann}. A system integration plan of the MOPS chip that includes powering and communication has been proposed with MOPS-HUB interface as the central unit. MOPS-HUB will act as a bidirectional interface between the new MOPS chips and the local control stations of the DCS.

\section{The new Monitoring of Pixel System Chip (MOPS)}
\label{sec:mops}
The DCS for the ATLAS ITK Pixel detector consists of three separate paths which are: \textbf{Diagnostics}, \textbf{Control \& Feedback} and \textbf{Safety} of all power supplies. The \textbf{Control \& Feedback} path implements the control and monitoring of the power supply and an independent monitoring of individual detector modules~\cite{CERN-LHCC-2017-021}. The monitoring of the detector modules will be done using new ADC  ASIC called MOPS. A CANopen based protocol is implemented in  it for data transfer. The new MOPS chip was developed to perform these tasks under high radiation exposure (up to \SI{500}{\mega\radian}). One MOPSchip can monitor the temperature and voltage of up to 16 detector modules in a serial power chain~\cite{Walsemann}. 

\comm{The MOPS chain then communicates to the main DCS computer over the CAN bus (see Fig.~\ref{fig:dcs_path}). 
\begin{figure}[htbp]
\centering
\includegraphics[width=0.8\textwidth]{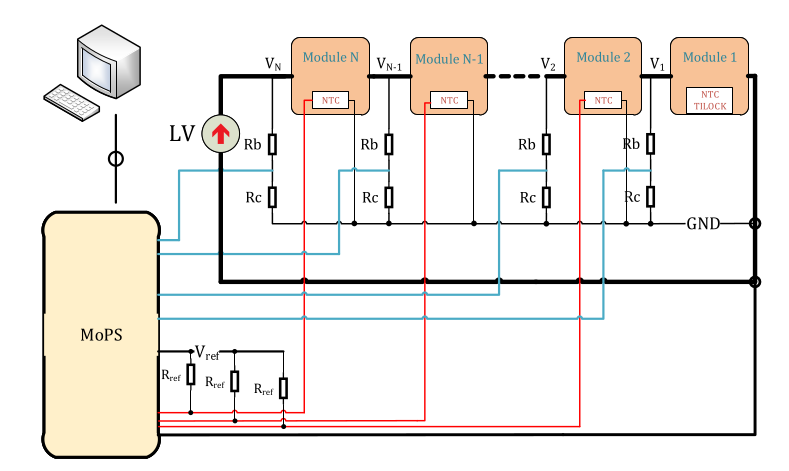}
\caption{\label{fig:dcs_path} Interconnection of the MOPS chip within the ITK Pixel detector for voltage and temperature measurements of the detector modules, based on~\cite{Walsemann}.}
\end{figure}. 
}
\section{Integration Plan of the MOPS Chip}
\label{sec:integration_mops}
MOPS-HUB is an FPGA based interface. Its main task is the aggregation of monitoring data between the~MOPS chips (installed at the vicinity of the detector modules) and the DCS computer. Beside that, MOPS-HUB will monitor information per CAN bus (voltage/current) and send it to the DCS computer as part of the data stream. In addition, it will have a full power control over the connected CAN buses.\newline 

\subsection{Interface to the Backend}
\label{sec:emci_emp}
\begin{figure}[ht]
\centering
\includegraphics[width=0.8\textwidth]{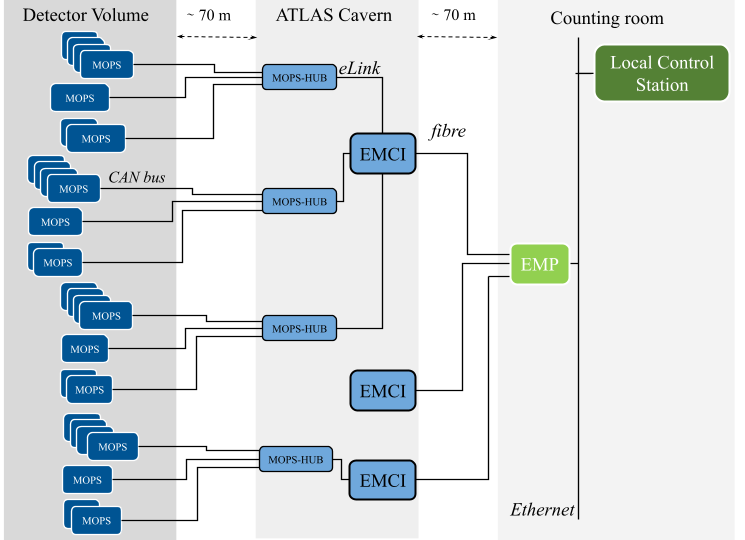}
\caption{\label{fig:construction} the complete MOPS-HUB network.}
\end{figure}

Depending on the configuration, the MOPS-HUB FPGA can interface simultaneously up to 16 CAN buses in the vicinity of the detector modules direction (see Fig.~\ref{fig:construction}). Each CAN bus is a serial control bus, which allows several CAN nodes on the same CAN bus (up to 4 MOPS). The MOPS-HUB itself will be placed in racks on the walls of the ATLAS cavern which is roughly \SI{70}{\m} away from the vicinity of the detector modules. The data collected from the CAN buses are aggregated through low power differential signals called elinks to the Embedded Monitoring and Control Interface (EMCI) which is also placed in racks on the walls of the ATLAS cavern~\cite{emci_emp}. The aim of the EMCI, is to work as as an further data aggregator between MOPS-HUB and the DCS. It combines all signals to one bidirectional channel and interfaces with an optical transceiver, which then transmits the data through an optical link to the Embedded Monitoring Processor (EMP)~\cite{emci_emp}. The EMP device will be placed in the counting room (see Fig.~\ref{fig:construction}). It will drive up to 12 EMCIs and interfaces them to the control network (Ethernet). Both ~EMCI and EMP are ATLAS standard DCS components. \newline
In addition, MOPS-HUB will distribute the power (VCAN-PSU), which is delivered from the main power supplies, onto the individual MOPS services. Where, the FPGA powering itself will be completely independent of any CAN bus powering.

    \subsection{Target FPGA}
\label{sec:targetfpga}
The FPGA chosen as MOPS-HUB core is the \textbf{XC7A200T} from Xilinx's Artix 7 Series~\cite{artix7_datasheet}. It features $215$,$360$ Logic Cells, 730 DSP slices and \SI{13140}{\kilo bit} of Block RAM. The design will use the 484-Pin FBG484 package and therefore have access to 285 user I/O pins~\cite{xilinx_7series_overview}. The FPGA clock is supplied by an external \SI{200}{\mega\hertz} active differential crystal. The exact part number of the FPGA is \textbf{XC7A200T-2FBG484I}.

\section{Firmware Description}
\label{sec:firmware_description}
\begin{figure}[ht]
	\centering
	\includegraphics[width=0.8\textwidth]{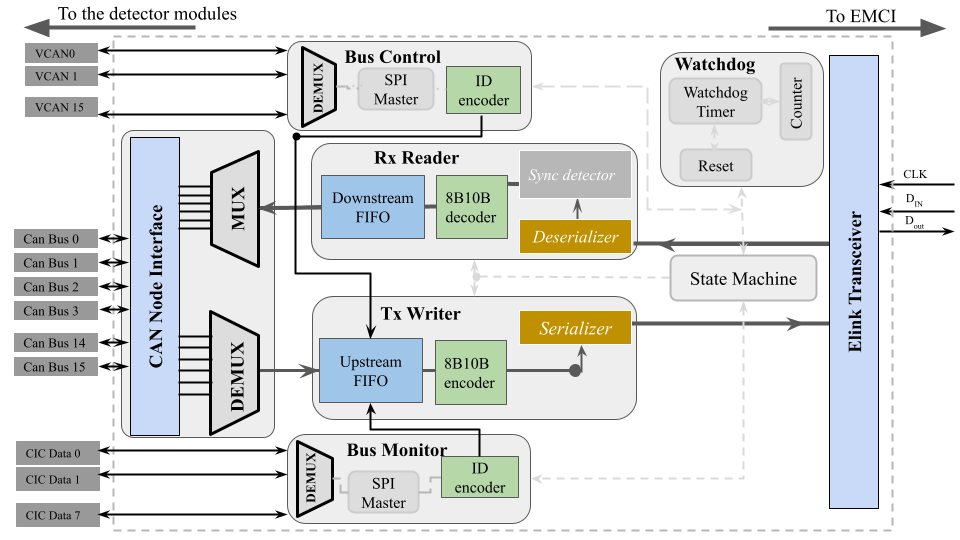}
	\caption{Firmware design of MOPS-HUB FPGA. The black bold lines represent data lines while the light gray lines represent the control signal from/to the top State Machine (FSM).}
	\label{fig:mopshub_firmware}
\end{figure}

The Firmware design of the MOPS-HUB FPGA consists of several logic blocks that contain the actual configurable logic. Additional resources of the FPGA are also used for clock management, dedicated memory, I/O and digital signal processing. Fig.~\ref{fig:mopshub_firmware} presents a block diagram of the MOPS-HUB FPGA firmware.\newline
As seen in Fig.~\ref{fig:mopshub_firmware}, the firmware design consists of eight components:
\begin{enumerate}
\item \textbf{CAN Node Interface}: collects the signals from all the inputs and send it to the next step for processing. It includes all the blocks needed for CAN communications (e.g.~\textbf{CAN controllers}).

\item \textbf{The TX Writer}: instantiates the memory blocks (e.g.~\textbf{Upstream FIFO}) module and the \textbf{8B10B encoder}). The data coming from the 8B10B encoder will be multiplexed out on \SI{2}{bit} port before transmission on the elink port. In this stage the \textbf{8B10B encoder} will encode the data so that a DC-Balanced code will be transmitted over each elink data line. 

\item \textbf{The RX Reader}: instantiates the memory blocks (e.g.~\textbf{Downstream FIFO}) and the \textbf{8B10B decoder}). The data in this block is synchronized, deserialized and aligned to 8B10B symbols once received from the elink side.   

\item \textbf{The elink Transceiver}: provides all the necessary building blocks needed for transmission and serialization over to the elinks. All the sub-modules included (e.g OSERDESE2, ISERDESE2, IBUFDS and OBUFDS) are supported from the manufacturer (e.g. Xilinx).

\item \textbf{Bus Control}: enables/disables the VCAN of the connected buses using SPI communication with the help of an external ADC. The module is designed in a way that the status of VCAN is preserved. Where, even if the FPGA is power cycled for any reason, e.g. system crash due to errors in the communication, the status of VCAN will be unchanged. 

\item \textbf{Bus Monitoring}: transfers the monitoring data read by external ADCs to the data stream over elinks. Where, the monitoring information per CAN bus (voltage/current of VCAN) will be acquired via an SPI bus.  

\item \textbf{State Machine (FSM)}: brings up the system in a working state , synchronizes the signal and manage processes between all the design components. Beside that, it will supervise the data transfer between CAN buses and elinks.

\item \textbf{Watchdog}: monitors the status of the top FSM after powering up. Once the FSM hangs for about 1-2 seconds due to errors, a timeout signal (generated by the \textbf{Watchdog Timer}) will be generated and an automatic recovery from interruptions will be activated. This automatic recovery will bring the system into a safe state.
\end{enumerate}

\section{Re-configuration}
\label{sec:error_handle}
To protect the system from firmware malfunction, the \textbf{Watchdog} module described in Sec.~\ref{sec:firmware_description} is developed to recover the system without requiring an external reset or power-cycle.\newline The effects related to S-RAM based FPGAs (e.g. flip-flops and memories) will be considered using \textbf{Scrubbing method} where the configuration memory of the used FPGA will be fully or partially re-written while the user design is running without any disruption. This will be done using an external, radiation tolerant memory that performs the boot-up of the FPGA and write the new configuration to it during operation.\newline
Another additional option is a re-configuration technique which enables updating the firmware from a remote location via elinks. In case that the MOPS-HUB FPGA does not respond anymore it must be reset and reconfigure itself.

\section{Radiation Tolerance}
\label{sec:radiation-tolerance}
As mentioned in Sec.~\ref{sec:integration_mops}, the MOPS-HUB will be operating in a radiation environment (PP3). This will pose special requirements on the design and firmware. Tab.~\ref{tab:rad_tolerance} shows the expected radiation at racks on the walls of the ATLAS cavern  based on the radiation background simulation results in the ATLAS hall Using FLUKA (The simulation results are provided by ATLAS collaboration). The results shown in Tab.~\ref{tab:rad_tolerance} include no safety factor. However, a safety factor of 1.5 will be considered as the minimum level of the radiation tolerance for the MOPS-HUB.\newline
The The total ionizing dose (TID) irradiation test of the chosen MOPSHUB-FPGA (\textbf{XC7A200T}) shows that the \textbf{XC7A200T} FPGA  can survive till up to \SI{550}{\kilo\radian}~\cite{Hu2019}.
\begin{table}[htbp]
\centering
\caption{\label{tab:rad_tolerance} Expected radiation at racks on the walls of the ATLAS cavern based on the simulation results using FLUKA in the ATLAS hall. No safety factors have been applied (The simulation results are provided by ATLAS collaboration).}
\smallskip
\begin{tabular}{|l|c|}
\hline
 Radiation Type &Expected Dose\\
\hline
TID       & \SI{30}{\gray}\\
Neutron fluence & \SI{5e11}{N_{eq}\per\cm^2}\\
Hadron fluence (>\SI{20}{\mega\eV})  &  \SI{2e-7}{\cm^2\per{pp}}\\
\hline
\end{tabular}
\end{table}
\subsection{Triple Modular Redundancy}
\label{sec:tmr}
TMR technique will be applied to the design in order to minimize radiation induced errors and enhance data reliability. In this case, If an SEU affects one of the three aforementioned hardware instances, the output will be based on the majority voting of all three outputs, thus the error will not propagate throughout the design. The feedback of the output allows for the voting mechanism to restore the proper state on the next clock cycle. By the nature of triplication, two errors cannot be fixed by the voting mechanism but is signalized by an error signal. Three errors will result in a failure. To minimize hardware utilization, TMR implementation will be carefully implemented by covering all critical parts of the firmware. To stay within the hardware constraints of the FPGA, only memory elements and no logic will be triplicated. 

\subsection{Radiation testing}
\label{sec:Radiation_testing}
In order to make sure that the design is resistant to damage or malfunction(transients or
permanent) caused by high levels of ionizing radiation, the full MOPS-HUB system will be tested against all the expected radiation effects mentioned in Tab.~\ref{tab:rad_tolerance}. This includes~TID , neutron or hadron displacement damage which causes SEUs.

\subsection*{TID measurements}
In order to perform TID dose studies, the electronic devices of MOPS-HUB will be irradiated using a Co-60 source (e.g Co-60 at ENEA-Casaccia Research Center~\cite{ENEA}). 

\subsection*{Sensitivity against SEUs}
To validate the implemented actions (e.g.~TMR) against SEUs, the error rate of SEUs will be measured at a high radiation test (Heavy Ion  or a Proton) facility (e.g. U.C.L\footnote{Université Catholique de Louvain}~\cite{ucl_ref} or CHARM \footnote{Cern High Energy AcceleRator Mixed-Field}~\cite{charm_ref},...etc). 

\section{Summary and Outlook}
A new interface called MOPS-HUB for data aggregation between the new monitoring chip (MOPS) and the DCS has been introduced. The central unit of the MOPS-HUB is the MOPS-HUB FPGA which will supervise the data transfer between MOPS and the DCS and manage processes between all the hardware components. The new interface will be operating in a radiation environment (racks on the walls of the ATLAS cavern) which poses special requirements on the design and firmware. This will require special testing in a radiation environment with the same environment characteristics as the final location to ensure the system reliability. 
\comm{
\appendix
\section{Some title}
Please always give a title also for appendices.
}
\comm{
\acknowledgments
This work was supported by the Federal Ministry of Education and Research, Germany.
We thank x (UPenn), y (CERN) and z (University of Bonn) for providing IP blocks. Our thanks go to all colleagues from the ATLAS community helping us
with the design. Special thanks go to x and x (University of Bonn) for the
wirebonding.
\paragraph{Note added.} This is also a good position for notes added
after the paper has been written.}


\end{document}